\documentclass[runningheads,orivec]{llncs}
\usepackage[utf8]{inputenc}
\usepackage[english]{babel}

\usepackage{graphicx}
\usepackage{transparent} 

\usepackage{tikz} 
\usetikzlibrary{automata, positioning, arrows, calc}
\usepackage{float}
\graphicspath{{Images/}} 

\usepackage{silence} 
\usepackage[font=scriptsize]{subfig} 
\usepackage[normalem]{ulem} 

\usepackage{amssymb}
\usepackage{amsmath}
\usepackage[overload]{empheq}  

\usepackage{tabularx}
\usepackage{longtable} 
\usepackage{colortbl}

\usepackage[ruled, vlined, linesnumbered]{algorithm2e}

\usepackage[colorlinks=true,linkcolor=black,anchorcolor=black,citecolor=black,filecolor=black,menucolor=black,runcolor=black,urlcolor=black]{hyperref} 

\usepackage{appendix}

\usepackage{enumitem}

\usepackage{lineno}

\usepackage{ifthen}
\newboolean{compileExtendedVersion}
\setboolean{compileExtendedVersion}{true}

\usepackage{orcidlink}

\usepackage{pdfpages}

\usepackage{comment} 
\usepackage{tcolorbox} 
\usepackage{pgfplots}
\pgfplotsset{compat=1.18}
\usepackage[textwidth=4cm,textsize=footnotesize]{todonotes}

\newcommand{\bea}{\begin{eqnarray}} 
\newcommand{\eea}{\end{eqnarray}}

\newcommand{\ite}{\ifthenelse{\boolean{compileExtendedVersion}}}

\newcommand{\uwa}{\mathcal{W}_{\mathcal{A},w}}
\newcommand{\thetaWeight}{w_{\Theta}}
\newcommand{\tileWeight}{w_{\mathcal{T}}}
\newcommand{\Aflat}{\mathcal{A}_{\mathit{flat}}}

\AtBeginDocument{

}

\newcommand{\toolname}{TABEC}

\begin{document}
    
\title{Random Testing of Model Checkers for Timed Automata with Automated Oracle Generation
\thanks{The experimental results presented in this paper are reproducible using the artifact available at: 
\href{https://doi.org/10.5281/zenodo.14871008}{10.5281/zenodo.14871008}.}}{}


 
  \author{Andrea~Manini\orcidID{0009-0009-6103-9244}
    \and 
    Matteo~Rossi\orcidID{0000-0002-9193-9560}
    \and 
    Pierluigi~{San~Pietro}\orcidID{0000-0002-2437-8716}
}
    \authorrunning{A. Manini \and M. Rossi\and P. {San Pietro}} 
 \institute{Politecnico di Milano\\ 
     \email{firstname.lastname@polimi.it}}
 \titlerunning{Random Testing of Model Checkers with Oracle Generation}
    
    \maketitle 
    
    
    \begin{abstract}
     A key challenge in formal verification, particularly in Model Checking, is ensuring the correctness of the verification tools. 
Erroneous results on complex models can be difficult to detect, yet a high level of confidence in the outcome is expected.  
Indeed, these tools are frequently novel and may not have been thoroughly tested.
When standard benchmarks may be insufficient or unavailable, random test case generation offers a promising approach. 
To scale up, random testing requires comparing actual versus expected results, i.e., solving the oracle problem.

To address this challenge, this work introduces a novel theoretical framework based on a modular variant of Timed Automata (TA), called Tiled Timed Automata (TTA), for testing  model checkers operating with variations of TA, by building oracles based on Weighted Automata.
The framework is initially applied to verify model checkers solving the emptiness problem for Parametric TA and it is validated, 
in this specific scenario, by our tool, \toolname{}, which randomly generates tests predicting their expected outcome through automated oracle generation. 
Furthermore, the general nature of TTA facilitates the framework adaptation to model checkers solving other decidable problems on TA, as detailed for the minimum-cost reachability problem of Priced TA.
    \end{abstract}

\section{Introduction}
\label{sec:introduction}
Ensuring the correctness of systems subject to vital temporal constraints is crucial to prevent considerable economic or human losses.
To address this challenge, various formal techniques have been developed to rigorously verify such systems.
Due to the intricacy and difficulty of correctness proofs based on mathematical logic, automated
verification techniques, such as Model Checking, are now central to formal verification.
Model Checking relies on a formal model of the system under study to check its behavior against relevant formal properties. 

Timed Automata (TA) are a popular mathematical formalism used to model concurrent and real-time {systems}, which are among the most difficult to develop and test.
Since the primary purpose of  verification tools for TA is to ensure the behavioral correctness of modeled systems,
they must be
as error-free as possible.
However, ensuring the complete absence of errors in verification tools is challenging, since their intrinsic complexity {hinders the rigorous proof of their correctness}.
Model checkers are software artifacts, hence traditional software engineering techniques for detecting errors---e.g., testing~\cite{pytesting}---can be employed.

Testing formal verification tools usually consists in running a set of standard benchmarks, whose results are known and thus can be checked against the actual tool output. 
However, the number of benchmarks is often limited and they may fail to cover new features of the specification language if the latter is extended.
{The original motivation of this work was indeed introducing a new framework for rigorously testing the correctness of model checkers working with variations of TA.
We developed a tool implementing this framework using a particular class of Parametric TA (PTA).}
A well-established way
to be confident of the correctness both of the technique and of the tool is to run a very large set of tests. 
Hence, {in our implementation, we validated the correctness of the framework and tool by applying} random testing techniques~\cite{randomAgrawal}, often used in software engineering for evaluating the robustness and performance of programs.

{Automatically verifying the accuracy of testing results, particularly for ran\-dom\-ly-generated, non-trivial test cases, is a daunting task, known as the \emph{oracle testing problem}~\cite{howden}, still a significant research topic~\cite{testingsurvey,stfm,toward}.}
Contemporary research initiatives, particularly those aligned with formal methods, have dedicated significant attention to the problem of deriving test oracles from specifications expressed in different kinds of temporal logics~\cite{generatingOracles,oraclesForChecking,theTemporalRover,onTheConstruction,specification}.
{Examples building TA-based oracles can be found in~\cite{repairingTimed,automaticGeneration}.}
To the best of our knowledge, no 
existing techniques 
automatically generate test oracles for TA-based tools. 
The recent work~\cite{mutation} applies mutation testing to networks of TA, implementing new mutation operators and evaluating their effectiveness with Uppaal~\cite{lpw97}.

{The idea behind the test oracle generation underpinning our framework} consists in creating a set of \emph{tiles}, i.e.,
TA whose behavior can be easily characterized by inspection, and then combining them in a 
randomly-generated {automaton}, called Tiled TA (TTA). 
A TTA can then be ``flattened'' into a TA analyzed by our verification tool, called
\toolname{} (Timed Automata Builder and Emptiness Checker)~\cite{tabec}.
By abstracting a TTA as a Weighted Automaton, it is possible to {generate an oracle predicting} a priori the correct result of the verification, e.g., the range of values (if any) admissible for the {parameter (when considering PTA tiles)}.
Although TABEC has been implemented with reference to PTA, the theoretical concepts presented in this paper are general and can be easily adapted to other variants of TA,
{as outlined for illustrative purposes with Priced TA.}

The main contribution of this work is two-fold:
(i) {we introduce a novel theoretical framework based on TTA to facilitate the testing of model checkers for variations of TA.
First, by using a decidable variant of PTA, we show how to use our framework to verify model checkers solving the language emptiness problem for PTA; then, we outline how to adapt the framework to the minimum-cost reachability problem of Priced TA;}
(ii) we developed \toolname{}
to validate our theoretical results {with respect to the emptiness problem}.
By generating test oracles, \toolname{} predicts the range of values (if any) for the unknown parameter.

The organization of this paper is as follows: \autoref{sec:background} provides the theoretical foundations for this work.
\autoref{sec:emptiness} presents an algorithm to solve the language emptiness problem of the considered decidable PTA variant. 
\autoref{sec:forecasting} {derives the theoretical framework for model checkers solving the language emptiness problem of PTA}. 
\autoref{sec:tester} details experimental results. 
{\autoref{sec:pta} adapts the framework to Priced TA.}
\autoref{sec:future} concludes and outlines future developments.


\section{Theoretical background}
\label{sec:background}
\subsubsection{(Parametric) Timed Automata}
Let $\Sigma$ be a finite alphabet. 
An $\omega$\emph{-language} is the set of infinite words $\sigma = a_{0}a_{1}a_{2}\dots$ defined over $\Sigma$.
A \emph{time sequence} $\tau = \tau_{0} \tau_{1} \tau_{2}\dots$ is an infinite sequence  of non-negative real numbers
such that, for all $i \geqslant 0$, it holds that $\tau_{i} < \tau_{i+1}$ (strict monotonicity) and 
for all $t \in \mathbb{R}_{> 0}$ there is $i \geqslant 0$ such that $\tau_{i} > t$ holds (progression).
A \emph{timed }$\omega$\emph{-word} (sometimes simply called \emph{timed word}) is a pair $(\sigma, \tau)$, where $\sigma$ is an infinite word and $\tau$ is a time sequence.
%
Timed Automata are defined as follows~\cite{alur}:
\begin{definition}[Timed Automaton]
\label{def:tadef}
A \emph{Timed Automaton} $\mathcal{A}$ (shortened as TA) is a tuple $\mathcal{A} = (\Sigma, Q, q_{0}, B, X, T)$, where
$\Sigma$ is a finite input alphabet,
$Q$ is a finite set of locations,  
$q_{0}$ is the initial location,
$B \subseteq Q$ is a subset of locations, called accepting locations,
$X$ is a finite set of clocks, and
$T \subseteq Q \times Q \times \Gamma(X) \times \Sigma \times 2^X$ is a transition relation.
\end{definition}


The input alphabet $\Sigma$ does not affect the results presented in this paper, hence in the following we assume it to be the singleton {$\Sigma = \{ a \}$} or just omit it.

Clocks are special variables that can only be reset or checked against an integer constant. 
Their value grows linearly in time until a reset occurs.
When a clock is reset, its value becomes 0, from which it starts increasing again. 
The set $\Gamma(X)$ contains \emph{clock guards}---i.e., predicates {over clocks} that must be satisfied for a transition to fire; 
they are specified according to the following grammar: $\gamma := x < c \; | \; x = c \; | \; \lnot \gamma \; | \; \gamma \land \gamma$, where $\gamma \in \Gamma(X)$, $x \in X$, and $c \in \mathbb{N}$. 
The powerset $2^{X}$ 
indicates
that a transition may reset a {subset} of the clocks.

A pair $(q, v)$ is called a \emph{configuration} for a TA, where $q \in Q$ is a location and $v : X \rightarrow \mathbb{R}_{\geqslant 0}$ is called a \emph{clock valuation}. 
A \emph{run} $\rho$ over a timed  word $(\sigma, \tau)$  is an infinite sequence 
$\rho = (q_{0}, v_{0})(q_{1}, v_{1})\dots $ 
of configurations starting in location $q_0$ 
such that $v_0(x) = 0$ for all $x \in X$ and, if $(q_{i}, v_{i})(q_{i+1}, v_{i+1})$ is a pair of consecutive configurations in the run, 
then there is a transition $(q_i, q_{i+1}, \gamma_i, \sigma(i), Y_i)\in T$, with $v_i$ satisfying $\gamma_i$, and $v_{i+1}$ is such that the clocks in $Y_i$ are 0, while the others are incremented by $\tau_{i+1}-\tau_{i}$.
Let \emph{inf}$(\rho)$ denote the set of locations visited infinitely often in a run.
A run $\rho$ is \emph{accepting} according to a B\"{u}chi acceptance condition if, and only if, \emph{inf}$(\rho) \cap B \neq \varnothing$ holds---i.e., $\rho$ enters one or more accepting locations infinitely often.
A timed $\omega$-word is accepted if it has an accepting run. 

TA can be enriched by adding a set $P$ of parameters to \autoref{def:tadef}, obtaining Parametric TA (shortened as PTA)~\cite{parres}.
Parameters enable flexible system representation by describing characteristics unknown at modeling time.
The value a parameter can assume is determined by a mapping $\mathcal{I} : P \rightarrow \mathbb{R}$, where $\mathcal{I}$ is called a \emph{parameter valuation}.
Clock guards in PTA include 
the following additional rules in their grammar: $x < \mu$ and $x = \mu$, where $\mu \in P$.
The notion of run for a PTA can be trivially extended to consider parameter valuations.
We call a run $\rho$ of a PTA, with parameter valuation $\mathcal{I}$, a \emph{parametric run}.

This work focuses on a particular version of TA, defined as follows~\cite{partimeforArXiv}:
\begin{definition}[Non-resetting test TA]
\label{def:nrttadef}
Let $\mathcal{A} = (\Sigma, Q, q_{0}, B, X, T)$ be a TA. For each transition  $u \in T$ of $\mathcal{A}$ of the form $u = (q_{u}, q'_{u}, \gamma_{u}, a_{u}, Y_{u})$, let $X(\gamma_{u})$ be the set of clocks that appear in $\gamma_{u}$. Then, $\mathcal{A}$ is called a \emph{non-resetting test Timed Automaton} (shortened as nrtTA) if,
for all $u \in T$, $X(\gamma_{u}) \cap Y_{u} = \varnothing$.
\end{definition}
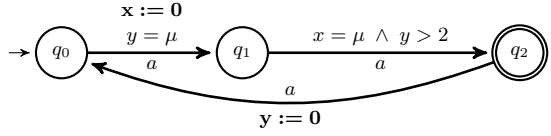
\begin{figure}[t]
\newcommand{\creset}{\textcolor{black}}
\newcommand{\cguard}{\textcolor{black}}
\begin{center}
\begin{tikzpicture}[
	->,
	>=stealth',
	shorten >=2pt, 
	auto,
	scale=0.85,
    transform shape, 
    align=center,
    state/.style={thick, circle, draw, minimum size=0.8cm}
    ] 
    
	\node[state, initial, initial text={}] (s0) {$q_0$};
	\node[state, right = 2cm of s0] (s1) {$q_1$}; 
	\node[state, accepting, right = 3.5cm of s1] (s2) {$q_2$}; 
		    
	\draw [line width=0.35mm]
	(s0) edge node{\creset{$\mathbf{x \boldsymbol{:=} 0}$} \\ \cguard{$y = \mu$}} node[below]{$a$}(s1)
	(s1) edge node{\cguard{$x = \mu \; \land \; y > 2$}} node[below]{$a$}(s2)
	(s2) [out=-160,in=-20] edge node[below]{\creset{$\mathbf{y \boldsymbol{:=} 0}$}} node[above]{$a$}(s0);
\end{tikzpicture}
\end{center}
\caption[Example of PnrtTA]{Example of PnrtTA, where $\Sigma = \{a\}$, $X = \{x, y\}$, and $P = \{\mu\}$. Clock resets are represented in bold, while clock guards are represented in italic. The initial location is $q_0$, while the (only) accepting location is $q_2$.}
\label{fig:pta1}
\end{figure}

Basically, in an nrtTA, a clock cannot be used both in guards and resets within the same transition.
\autoref{def:nrttadef} can trivially be extended
by introducing a set $P$ of parameters, obtaining \emph{Parametric nrtTA} (shortened as PnrtTA).
In the remainder of this work, we focus on PnrtTA having two clocks and only one parameter, and refer to this variant simply as PnrtTA.
An example of a PnrtTA satisfying these requirements is given in \autoref{fig:pta1}.
This restriction enables the application of the theoretical results presented below.

\subsubsection{Decidability of PnrtTA}
In the recent work~\cite{partimeforArXiv}, the emptiness problem was proven to be decidable for PnrtTA.
%
The result is summarized as follows:

\begin{theorem}
\label{thm:parametric}
Let $\mathcal{A}$ be a PnrtTA and $C$ be the largest constant appearing in the guards of $\mathcal{A}$. Then:
~\begin{enumerate}
\item There exists a constant value $\Xi > 2C$ such that, $\forall \bar{\mu} \in \mathbb{R}: \bar{\mu} > 2C$, with $\bar{\mu} \neq \Xi$, if there is a parametric run $\rho$ for $\mathcal{A}$ 
having a parameter valuation equal to $\bar{\mu}$, then there is also a parametric run $\hat{\rho}$ for $\mathcal{A}$
 having a parameter valuation equal to $\Xi$.
\item There exists a constant value $0 < \alpha < \frac{1}{2}$ such that, $\forall n \in \mathbb{N} : 0 < n < 4C$, if there is a parametric run $\rho$ for $\mathcal{A}$
 having a parameter valuation equal to $\bar{\mu}$, with $\frac{n}{2} < \bar{\mu} < \frac{n + 1}{2}$, then there is also a parametric run $\hat{\rho}$ for $\mathcal{A}$ 
 over a timed word $(\sigma, \hat{\tau})$ 
 having a parameter valuation equal to $\hat{\mathcal{I}}(\mu) = \hat{\mu} = \frac{n}{2} + \alpha$.

\end{enumerate}
\end{theorem}

The two values $\Xi$ and $\alpha$ can be easily computed as a function of $C$ and of the number of locations of $\mathcal A$ (denoted as $\lvert Q \rvert$)~\cite{partimeforArXiv}.
\footnote{More precisely, $\Xi$ is any value greater than $1 + C ( 1 + \lvert Q \rvert )$,
while $\alpha$ is required to be any value less than 
$\frac{1}{4(1 + C \cdot \mathit{max}(\lvert Q \rvert, 4C))}$.}

\subsubsection{Weighted Automata}
Weighted Automata are used in this paper to derive test oracles.
Weights over a semiring facilitate the definition of test oracles that operate on different mathematical entities.
The most appropriate semiring must be manually chosen to align with the specific operational context of the problem at hand.
We follow the definitions of~\cite[Chapter 4]{handbookautomata}, with minor differences. 


A \emph{semiring} $\breve{S} = (S,\oplus, \otimes, 0_S,1_S)$ is an algebraic structure such that:  
(i) $S$ is a set, (ii) $\oplus$ (called sum) and  $\otimes$ (called product) are associative binary operations over $S$, (iii) $\oplus$ is commutative, (iv) $0_S$ is the identity of  $\oplus$, (v)
$1_S$ is the identity of $\otimes$, (vi) $\otimes$ distributes over $\oplus$ both from the left and the right, and (vii) $0_S$ is an annihilator for $\otimes$ (i.e., for all $s \in S$, $s \otimes 0_S = 0_S \otimes s = 0_S$).

\begin{example}\label{ex:semirings1}
For $k \geqslant 1$, let $\mathcal{B}_k$ be the set $\{0,1\}^k$ of Boolean words of length $k$, and $\mathit{OR}$, $\mathit{AND}$ be the corresponding bitwise operations.
Then, $W^{k-\text{bit}} = (\mathcal{B}_k, \mathit{OR}, \mathit{AND}, 0^k,1^k)$ is a semiring, since {(i)} $\mathit{OR}$, $\mathit{AND}$ have the required properties over Boolean words, {(ii)} $0^k$, $1^k$ are the identity for the $\mathit{OR}$ and $\mathit{AND}$ respectively, and {(iii)} $s \, \mathit{AND} \, 0^k~=~0^k \, \mathit{AND} \, s~=~0^k$ for every $s \in \mathcal{B}_k$.
\end{example}
        
\begin{example}\label{ex:semirings2}
Another example of semiring is $W^{\text{price}}= (\mathbb{N}\cup \{+\infty\}, \mathit{min}, +, +\infty,0)$, often called the tropical, or min-plus, semiring. 
In this case, weights represent costs to be summed and minimized; the value $+\infty$ is the identity of the $min$ operation since, e.g.,  $min(x,+\infty)= x$ for every $x \in \mathbb{N}$. 
\end{example}

\begin{definition}[Weighted Automaton]
\label{def:weightedta}
A \emph{Weighted Automaton} $\mathcal{W}$ {over a semiring $(S,\oplus, \otimes, 0_S,1_S)$} is a tuple $\mathcal{W}=(\Sigma, Q, q_0, F, w)$, where $\Sigma$ is a finite input alphabet, $Q$ is a finite set of states, $q_0\in Q$ is the initial state, ${F \subseteq Q}$ is a set of final states, and $w = w_T \cup w_F$ is a weight function, where $w_T: Q \times \Sigma \times Q \to S$ is the transition weight function and $w_F:F \to S$ is the final weight function.
\end{definition}

 A \emph{path} $P$ of $\mathcal{W}$ is a \emph{finite} sequence $q_0 y_1 q_1 y_2\dots y_n q_n$, for $n \geqslant 0$,
where $q_i \in Q$ for all $i \geqslant 0$, $q_0$ is initial, and $y_i \in \Sigma$. 
The \emph{label} of $P$ is $l(P)= y_1 \dots y_n$.
It is possible to extend the weight function $w$ to paths as follows: given a finite path  $P = {q}_0 y_1 {q}_1 y_2 \dots y_{n} {q}_n$, its weight is $0_S$ if $q_n$ is not final, otherwise it is: 
\begin{equation*}\label{eq:prod}
    w(P)= \left( \prod_{1 \leqslant i \leqslant n} w_T(q_{i-1}, y_i, q_i) \right) \otimes w_F(q_n)
\end{equation*}
where $\prod$ uses the $\otimes$ operation for the product. 

The \emph{behavior} of $\mathcal{M}$ is the mapping $w_l : \Sigma^* \to S$ defined for all $y \in \Sigma^*$ as: 
\begin{equation*}\label{eq:sum}
    w_l(y)= \sum_{P:l(P)=y} w(P),
\end{equation*} 
where $\sum$ uses the $\oplus$ operation for the sum. 

For example, in a Weighted Automaton $\mathcal{W}$ defined over the min-plus semiring of \autoref{ex:semirings2}, a weight may represent a cost (a price) to pay to traverse a transition of $\mathcal{W}$.
Costs are summed up along a path, {and the cost of a given word is defined as the minimum cost among all paths labeled with that word.}
Paths ending in a non-final state have weight $+\infty$.
The behavior of $\mathcal{W}$ is just the mapping defining the cost of each word. 

\section{Checking PnrtTA emptiness}
\label{sec:sec4}\label{sec:emptiness}
We developed and implemented an algorithm based on \autoref{thm:parametric} to determine, given a PnrtTA $\mathcal{A}$, whether its language is empty.
The key idea consists in replacing the parameter with suitable constant values, so that checking the emptiness of $\mathcal{A}$ is reduced to checking the emptiness of several TA without parameters.

A pseudocode version of this algorithm is reported in \autoref{alg:mainalgo}, called \hyperref[alg:mainalgo]{EmpCheck} (we also refer to it as EmpCheckFast when \emph{fstFlag} is true).
%
Given a PnrtTA $\mathcal{A}$ as input, \hyperref[alg:mainalgo]{EmpCheck} can find all parameter values that can lead to a B\"uchi acceptance condition in $\mathcal{A}$. 
In particular, line 2 checks the case in which $\mu > 2C$. 
A total of $4C + 1$ iterations are performed from line 6 to line 10 to check the case in which $\mu \leqslant 2C$ and $\mu$ is a multiple of $\frac{1}{2}$ holds. 
Lastly, if the condition on line 11 is not satisfied, a total of $4C$ iterations are performed from line 13 to line 17 to check the case in which $\mu < 2C$ holds.
If any of the conditions on lines 3, 8, or 15 are satisfied, the algorithm halts immediately.

The \texttt{CheckNonParEmptiness($\mathcal{A}, \bar{\mu}$)}
procedure substitutes each parameter occurrence in $\mathcal{A}$ with a specific constant value $\bar{\mu}$ derived from~\autoref{thm:parametric}, depending on the considered case; this substitution yields a usual TA without parameters.
The resulting automaton {is then checked by the same procedure, leveraging} existing verification tools
capable of detecting B\"uchi acceptance conditions in TA.\begin{algorithm}[h]
    \caption{EmpCheck}
    \label{alg:mainalgo}
    
    \KwData{$\mathcal{A}$: a PnrtTA, \emph{fstFlag}: a Boolean value.}
    \KwResult{\emph{true} if the language of $\mathcal{A}$ is not empty, \emph{false} otherwise.}
    \Begin() {
   
    $\mathit{isAccepting}$ $\gets$ \texttt{CheckNonParEmptiness}$(\mathcal{A}, \Xi)$\;
    
    \If{$(\mathit{fstFlag} \land \mathit{isAccepting}) = \mathit{true}$}{
        \Return \emph{true}\;
    }
      $n \gets 0$\;
      
      \While{$n \leqslant 2C$} {
        \emph{isAccepting} $\gets$ \emph{isAccepting} $\lor$ \texttt{CheckNonParEmptiness}$(\mathcal{A}, n)$\;

        \If{$(\mathit{fstFlag} \land \mathit{isAccepting}) = \mathit{true}$}{
        \Return \emph{true}\;
    }
        $n \gets n + \frac{1}{2}$\;
      } 
      \If{$\mathit{isAccepting} = \mathit{false}$}{
        $n \gets 0$\;
        
        \While{$n < 4C$} {
          \emph{isAccepting} $\gets$ \emph{isAccepting} $\lor$ \texttt{CheckNonParEmptiness}$(\mathcal{A}, \frac{n}{2} + \alpha)$\;

          \If{$(\mathit{fstFlag} \land \mathit{isAccepting}) = \mathit{true}$}{
        \Return \emph{true}\;
     } 
          $n \gets n + 1$\;
        }
        }
        \Return \emph{isAccepting}\;
    }
\end{algorithm}

Since the \texttt{CheckNonParEmptiness}$(\mathcal{A}, \bar{\mu})$ procedure is PSPACE-complete, it follows that both \hyperref[alg:mainalgo]{EmpCheck} and EmpCheckFast are PSPACE-complete as well.

\begin{figure}[t]
\centering
\includegraphics[width=\textwidth]{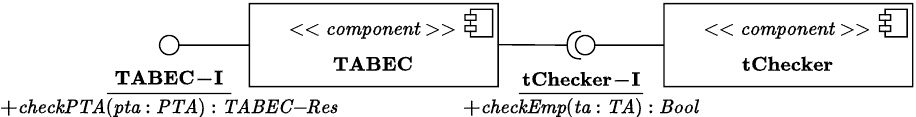}
\caption[Component diagram]{Component diagram illustrating the interaction between \toolname{} and tChecker. As highlighted by the \emph{tChecker-I} interface and its \emph{checkEmp()} method, tChecker is only responsible for verifying the emptiness of a given TA $\mathcal{A}$ and returning the result (either true or false) to \toolname{}. The \emph{\toolname{}-I} interface defines the functionalities provided by \toolname{}. The return type \emph{TABEC-Res} of the \emph{checkTA()} method represents a custom output type containing emptiness results, as well as possible values of the parameter leading to a B\"uchi acceptance condition in $\mathcal{A}$.}
\label{fig:component}
\end{figure}

The current implementation of TABEC uses tChecker~\cite{tcheckerpaper} as the verification engine for emptiness checking of TA.
The widely used tool Uppaal is not suitable for this task, as it does not support the detection of B\"uchi acceptance conditions in TA, unlike tChecker.
The component diagram of \autoref{fig:component} illustrates the interaction between \toolname{} and tChecker. 
The latter is responsible for analyzing the emptiness of a given TA and returning the results via the \emph{tChecker-I} interface.

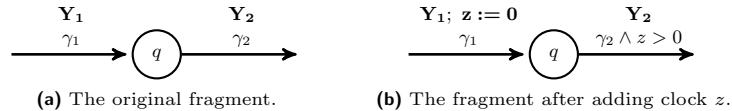
\begin{figure}[b]
	\newcommand{\creset}{\textcolor{black}}
	\newcommand{\cguard}{\textcolor{black}}
    \centering
    \subfloat[The original fragment.]{
    \begin{tikzpicture}[
	->,
	>=stealth',
	shorten >=2pt, 
	auto,
	scale=0.8,
    transform shape, 
    align=center,
    state/.style={thick, circle, draw, minimum size=0.8cm}
    ] 
    
->,
	>=stealth',
	shorten >=2pt, 
	auto,
	scale=0.8,
    transform shape, 
    align=center,
    state/.style={thick, circle, draw, minimum size=0.8cm}
    ] 
    
	\node[state] (s0) {$q$};
	\node[state, scale = 0.3, draw=white, right = 2cm of s0] (s1) {$ $}; 
	\node[state, scale = 0.3, draw=white, left = 2cm of s0] (s2) {$ $};
		    
	\draw [line width=0.35mm]
	(s0) edge node{\creset{$\mathbf{Y_2}$} \\ \cguard{$\gamma_2$}} (s1)
	(s2) edge node{\creset{$\mathbf{Y_1}$} \\ \cguard{$\gamma_1$}} (s0)
	
    ;    
    \end{tikzpicture}
    }
    \quad
    \subfloat[The fragment after adding clock $z$.]{
    \begin{tikzpicture}[
	->,
	>=stealth',
	shorten >=2pt, 
	auto,
	scale=0.8,
    transform shape, 
    align=center,
    state/.style={thick, circle, draw, minimum size=0.8cm}
    ] 
    ->,
	>=stealth',
	shorten >=2pt, 
	auto,
	scale=0.8,
    transform shape, 
    align=center,
    state/.style={thick, circle, draw, minimum size=0.8cm}
    ] 
    
	\node[state] (s0) {$q$};
	\node[state, scale = 1, draw=white, right = 2cm of s0] (s1) {$ $}; 
	\node[state, scale = 1, draw=white, left = 2cm of s0] (s2) {$ $};
		    
	\draw [line width=0.35mm]
	(s0) edge node{\creset{$\mathbf{Y_2}$} \\ \cguard{$\gamma_2 \land z > 0$}} (s1)
	(s2) edge node{\creset{$\mathbf{Y_1; ~ z \boldsymbol{:=} 0}$} \\ \cguard{$\gamma_1$}} (s0)
	
    ;
	    \end{tikzpicture}
    }
    \caption[Example of transformation.]{Illustrative transformation resulting from the addition of clock $z$ to a nrtTA fragment consisting of a single location with one incoming and one outgoing transition. 
    The symbols $Y_1, Y_2$ and $\gamma_1,\gamma_2$ denote, respectively, generic resets and guards.}
    \label{fig:transform}
\end{figure}

There is a slight difference between the nrtTA model used in this work and the one that is given to tChecker by \toolname{}. 
In tChecker, time sequences are only weakly monotonic (i.e., time may also not advance between two transitions), while in the nrtTA model they are strictly monotonic. 
To force strict monotonicity, the set of clocks of a given nrtTA $\mathcal{A}$ is enriched with an additional clock $z$.
Since clock $z$ is never compared with constants (other than 0) or parameters, it does not affect the language accepted by $\mathcal{A}$,
{thus preserving emptiness results before and after its addition.}
The new clock $z$ is reset on every transition entering a location and introduces an additional condition $\gamma := z > 0$ {to be added to preexisting guards} on every transition exiting from a location. 
This process transforms $\mathcal{A}$ into an equivalent TA that no longer retains the non-resetting test property. 
The content of this paper is not affected, as this is only an implementation mechanism to achieve strict monotonicity in practice.
A graphical example of this transformation is given in \autoref{fig:transform}.

\section{Oracle Generation Framework}
\label{sec:sec5}\label{sec:forecasting}
{This section introduces tiles and Tiled TA, used to create the oracles needed to verify the correctness of model checkers for TA.
We begin by providing general definitions and then focus on the emptiness problem for Parametric TA.}

\subsubsection{Tiled Timed Automata}
Tiles are particular TA that can be considered as building blocks for creating more complex TA, having an effect which can be easily predicted.
The latter varies based on the nature of the considered tiles, as will be shown later for parametric tiles and in \autoref{sec:pta} for priced tiles.


\begin{definition}[Tile]
    \label{def:tiledef}
    A \emph{tile} $\mathcal{T}$ is a tuple $\mathcal{T} = (\Sigma, Q, B, X, T, \mathit{In}, \mathit{Out}, \mathcal{F})$, where $\Sigma$, $Q$, $B$, $X$, $T$ are as in \autoref{def:tadef} and where
    $\mathit{In} \subseteq Q$ is the set of \emph{input} locations,
    $\mathit{Out} \subseteq Q$ is the set of \emph{output} locations, $B$ may be empty, and $\mathcal{F}$ is a set containing the functions (called compatibility functions)
    $\mathcal{I}_{\mathcal{T}, \gamma} : \mathit{In} \rightarrow \Gamma(X)$,
    $\mathcal{I}_{\mathcal{T}, Y} : \mathit{In} \rightarrow 2^X$,
    $\mathcal{O}_{\mathcal{T}, \gamma} : \mathit{Out} \rightarrow \Gamma(X)$, and
    $\mathcal{O}_{\mathcal{T}, Y} : \mathit{Out} \rightarrow 2^X$.
\end{definition}


{Let $\mathcal{T}$ be a tile.}
When needed, we use $\mathcal{T}$ as subscript to separate the elements of different tiles, e.g., $\mathit{In}_{\mathcal{T}}$ and $\mathit{Out}_{\mathcal{T}}$ indicate, respectively, the input and output sets of $\mathcal{T}$.
If $\mathcal{T}$ is such that both $B_{\mathcal{T}} \neq \varnothing$ and $\mathit{Out}_{\mathcal{T}} = \varnothing$ hold, then $\mathcal{T}$ it is called an \emph{accepting tile}.
%
%
Input and output locations, along with compatibility functions, are used to connect tiles together according to the following relation:

\begin{definition}[Tile transition relation]
    \label{def:tilesequence}
    Given an alphabet $\Sigma$ and a set $\Theta$ of tiles, let 
    $X_{\Theta} = \bigcup_{\mathcal{T} \in \Theta}X_{\mathcal{T}}$ be
    the set of clocks of $\Theta$.
    A \emph{tile transition relation}  $\Upsilon$ is a set of tuples included in 
    $
    \bigcup_{\mathcal{T}, \mathcal{T}' \in \Theta} \mathit{Out}_{\mathcal{T}} \times \mathit{In}_{\mathcal{T}'} \times \Gamma(X_{\Theta}) \times \Sigma \times 2^{X_{\Theta}}
    $
    such that,
    for all transitions $(q, q', \gamma, a, Y) \in \Upsilon$,
    the assume-guarantee constraint $\mathcal{O}_{\mathcal{T}, \gamma}(q) \Rightarrow{}  \mathcal{I}_{\mathcal{T}', \gamma}(q')$ holds,
    $\gamma = \mathcal{O}_{\mathcal{T}, \gamma}(q)$, and 
    $Y= 
    \mathcal{O}_{\mathcal{T}, Y}(q)
    \cup
    \mathcal{I}_{\mathcal{T}', Y}(q')$.
\end{definition}

Since accepting tiles have no output locations, a tile transition relation does not contain transitions exiting from them.
\autoref{def:tilesequence} specifies how different tiles are connected together, enabling the construction of arbitrarily large TA:


\begin{definition}[Tiled TA]
\label{def:tiledTA}
    A \emph{Tiled Timed Automaton} $\mathcal{A}$ (shortened as TTA) is a tuple $\mathcal{A} = (\Theta, \mathcal{T}_{0},\Sigma, \mathbb{B}, X_{\Theta}, \Upsilon)$, where
    $\Theta$ is a set of tiles,
    $\mathcal{T}_{0} \in \Theta$ is the initial tile,
    $\Sigma$ is a finite input alphabet,
    $\mathbb{B} \subseteq \Theta$ is a set of accepting tiles,
    $X_{\Theta}$ is a set of clocks defined as $X_{\Theta} = \bigcup_{\mathcal{T} \in \Theta} X_{\mathcal{T}}$, and
    $\Upsilon$ is a tile transition relation
    defined over $\Sigma$ and $\Theta$ as in \autoref{def:tilesequence}.
\end{definition}

{A B\"uchi acceptance condition for TTA is introduced 
by requiring that an accepting tile is reached and that one of {its}
final locations 
is visited infinitely often.
A TTA cannot exit an accepting tile; this allows it to accept infinite words while visiting its tiles only finitely many times.}

The \emph{flattening} of a TTA $\mathcal{A}$ results in a TA, denoted by $\Aflat{}$, obtained by substituting each tile within $\mathcal{A}$ with its corresponding TA.
%
%
The $\omega$-language accepted by $\mathcal{A}$, denoted by $\mathcal{L}(\mathcal{A})$, is defined as the $\omega$-language accepted by $\Aflat{}$.
For all runs $\rho = (q_{0}, v_{0})(q_{1}, v_{1})\dots$ of $\Aflat{}$ it must hold that $q_0 \in \mathit{In}_{\mathcal{T}_0}$ and $v_0(x) = 0$ for all $x \in X_{\Theta}$.

\begin{figure}[tb]
\newcommand{\creset}{\textcolor{black}}
\newcommand{\cguard}{\textcolor{black}}
\begin{center}
\begin{tikzpicture}[
	->,
	>=stealth',
	shorten >=2pt, 
	auto,
	scale=1,
    transform shape, 
    align=center,
    state/.style={thick, rectangle, draw, minimum size=0.6cm}
    ] 
    
	\node[state, initial, initial text={}] (s0) {$\mathcal{T}_{0}$};
	\node[state, right = 2cm of s0] (s1) {$\mathcal{T}_{1}$}; 
	\node[state, right = 2cm of s1] (s2) {$\mathcal{T}_{2}$};
	\node[state, accepting, right = 4.5cm of s2] (s3) {$\mathcal{T}_{3}$};  
		    
	\draw [line width=0.35mm]
    (s2) edge node{
    \creset{$\mathbf{Y_{23} \boldsymbol{=} \mathcal{O}_{\mathcal{T}_{\text{2}}, \text{$Y$}}}(q)
    \cup
    \mathbf{
    \mathcal{I}_{\mathcal{T}_{\text{3}}, \text{$Y$}}}(q')$} \\ 
    \cguard{$\gamma_{23} =
    \mathcal{O}_{\mathcal{T}_2, \gamma}(q)$}} (s3)
    
    (s1) edge node{\creset{$\mathbf{Y_{10}}$} \ 
    \cguard{$\gamma_{10}$}} (s0)
    
    (s0) [out=20,in=160] edge node{\creset{$\mathbf{Y_{02}}$} \ 
    \cguard{$\gamma_{02}$}} (s2)
    
    (s2) [out=180,in=0] edge node{\creset{$\mathbf{Y_{21}}$} \
    \cguard{$\gamma_{21}$}} (s1)
    ;
\end{tikzpicture}
\end{center}
\caption[Example of TTA]{TTA where tile $\mathcal{T}_{0}$ is initial, tile $\mathcal{T}_{3}$ is accepting, location $q \in \mathit{Out}_{\mathcal{T}_2}$, and location $q' \in \mathit{In}_{\mathcal{T}_3}$.
Resets and guards of the tile transition relation are shown, respectively, in bold and italic;
on the transition from tile  $\mathcal{T}_2$ to tile $\mathcal{T}_3$ they are made explicit.}
\label{fig:tta1}
\end{figure}
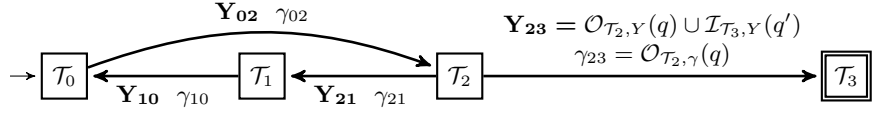

%
%
An example of TTA is shown in \autoref{fig:tta1} where, for a B\"uchi acceptance condition to hold, it suffices to reach tile  $\mathcal{T}_3$ starting from tile $\mathcal{T}_0$.

\subsubsection{Underlying Weighted Automaton}
Let $\mathcal{A} = (\Theta, \mathcal{T}_{0},\Sigma, \mathbb{B}, X_{\Theta}, \Upsilon)$ be a TTA.
We can associate each non-accepting tile $\mathcal{T} \in (\Theta \setminus \mathbb{B})$ with a weight function $\tileWeight{} : \mathit{Out}_{\mathcal{T}} \to S$, mapping each output location of $\mathcal{T}$ to a weight in the domain $S$ of an appropriate semiring.
The semiring and weights must be manually defined for each problem at hand. 
Since accepting tiles do not have output locations, we also define a weight function $w_{\mathbb{B}}: \mathbb{B}\to S$ associating a weight with each accepting tile.
The weight function over $\Theta$ is defined as $\thetaWeight{} =  w_{\mathbb{B}} \cup \bigcup_{\mathcal{T} \in (\Theta \setminus \mathbb{B})} \tileWeight{}$.
%
A particular
Weighted Automaton is derived
from $\mathcal{A}$ and $\thetaWeight{}$:

\begin{definition}[Underlying Weighted Automaton]
\label{def:underlyingWA}
    Let $\mathcal{A} = (\Theta, \mathcal{T}_{0},\Sigma, \mathbb{B}, $ $X_{\Theta}, \Upsilon)$ be a TTA and $\thetaWeight{}$ be a weight function over $\Theta$.
    The \emph{Underlying Weighted Automaton} of $\mathcal{A}$ is the Weighted Automaton
    $\uwa{} = (\Sigma, Q, q_0, F, w)$, where $Q = \Theta$, $q_0 = \mathcal{T}_0$, $F = \mathbb{B}$, and $w = w_T \cup w_F$ is such that, for all transitions $(q, q', \gamma, a, Y) \in \Upsilon$, where $q \in \mathit{Out}_{\mathcal{T}}$ and $q' \in \mathit{In}_{\mathcal{T}'}$, for $\mathcal{T} \in (\Theta \setminus \mathbb{B})$ and $\mathcal{T}' \in \Theta$,  
    $w_T(\mathcal{T}, a, \mathcal{T}') = \thetaWeight{}(q)$ holds
    and, for all $\mathcal{T} \in \mathbb{B}$, $w_F(\mathcal{T}) = \thetaWeight{}(\mathcal{T})$ holds.
\end{definition}
 
%
%
%
%
For instance, consider the TTA $\mathcal{A}$ of \autoref{fig:tta1}:  $\uwa{}$ is obtained by replacing guards and resets on each transition with weights. 
Then, the weight
of the path $P = \mathcal{T}_0 \mathcal{T}_2 \mathcal{T}_3$ in $\uwa{}$ 
is 
$w(P) = \thetaWeight{}(q_0) \otimes \thetaWeight{}(q) \otimes \thetaWeight{}({\mathcal{T}_3})$, where $q_0 \in \mathit{Out}_{\mathcal{T}_0}$.

Notice that an Underlying Weighted Automaton is always defined on finite runs, while a TTA only accepts infinite words.
Weights characterize the behavior of TTA, as shown later for PTTA and in \autoref{sec:pta} for Priced TA.


\subsubsection{Parametric TTA}
A \emph{parametric tile} is a non-resetting test tile with one parameter (denoted by $\mu$) and two clocks. 
{Parametric tiles can force $\mu$ to be inside one or more intervals.
Given the region characterization of~\cite{partimeforArXiv},
such intervals are of the form $\frac{a}{2} \sim_1 \mu \sim_2 \frac{b}{2}$, where $a \in \mathbb{N}$, $b \in (\mathbb{N} \cup \{+\infty\})$, $\sim_1, \sim_2 \; \in \{<, \leqslant\}$, and $a \leqslant b$}; for ease, the case $a=b$ defines a single point, still treated as an interval.
{A non-accepting parametric tile $\mathcal{T}$ can force $\mu$ inside one or more intervals
if there exists at least one path from $\mathit{In}_{\mathcal{T}}$ to $\mathit{Out}_{\mathcal{T}}$.
A set of intervals $\delta_q$, also called \emph{parameter set}, is forced for each reachable output location $q \in \mathit{Out}_{\mathcal{T}}$.
An accepting tile $\mathcal{T}$ has a parameter set $\delta_{\mathcal{T}}$, which is nonempty if there exists at least one infinite path from $\mathit{In}_{\mathcal{T}}$ visiting some location in $B_{\mathcal{T}}$ infinitely often}.
 
\begin{figure}[t]
\newcommand{\creset}{\textcolor{black}}
\newcommand{\cguard}{\textcolor{black}}
\begin{center}
\begin{tikzpicture}[
	->,
	>=stealth',
	shorten >=2pt, 
	auto,
	scale=0.8,
    transform shape, 
    align=center,
    state/.style={thick, circle, draw, minimum size=0.8cm}
    ] 
    
	\node[state, label=above:{in}] (s0) {$q_0$};
	\node[state, right = 2cm of s0] (s1) {$q_1$}; 
	\node[state, label=above:{out}, right=3cm of s1, yshift=1cm] (s2) {$q_2$};
    \node[state, label=above:{out}, right=3cm of s1, yshift=-1cm] (s3) {$q_3$};

    \node[state, scale = 0.3, draw=white, left = 3cm of s0] (s00) {$ $}; 
	\node[state, scale = 0.3, draw=white, right = 3cm of s2] (s01) {$ $};
    \node[state, scale = 0.3, draw=white, right = 3cm of s3] (s02) {$ $};

    

		    
	\draw [line width=0.35mm]
	(s0) edge node{\creset{$\mathbf{x \boldsymbol{:=} 0}$} \\ \cguard{$y = \mu$}} (s1)
	

    (s1) edge node[pos=0.8]{
    \cguard{$x = \mu \; \land \; y > 6$}
    } (s2)
    

    (s1) edge [below left] node[pos=0.8]{
    \cguard{$x = \mu \; \land \; y \leqslant 4$}
    } (s3)

    (s00) edge[dashed, lightgray] node[text=black] {
    $\mathcal{I}_{\mathcal{T}, Y}(q_0) = \{ y \}$
    \\
    $\mathcal{I}_{\mathcal{T}, \gamma}(q_0) = \mathit{true}$
    } (s0)

    (s2) edge[dashed, lightgray] node[text=black] {
    $\mathcal{O}_{\mathcal{T}, Y}(q_2) = \varnothing$
    \\
    $\mathcal{O}_{\mathcal{T}, \gamma}(q_2) = \text{``}y > 8\text{''}$
    } (s01)

    (s3) edge[dashed, lightgray] node[text=black] {
    $\mathcal{O}_{\mathcal{T}, Y}(q_3) = \varnothing$
    \\
    $\mathcal{O}_{\mathcal{T}, \gamma}(q_3) = \mathit{true}$
    } (s02)
	
    ;
\end{tikzpicture}
\end{center}

\caption{Tile forcing the intervals $\delta_{q_2} = \{ (3, + \infty) \}$ and $\delta_{q_3} = \{ (0, 2] \}$.
The tile is neither initial nor accepting.
Compatibility functions are reported over gray dashed arrows, which symbolically represent a tile transition relation.}
\label{fig:pta12short}
\end{figure}
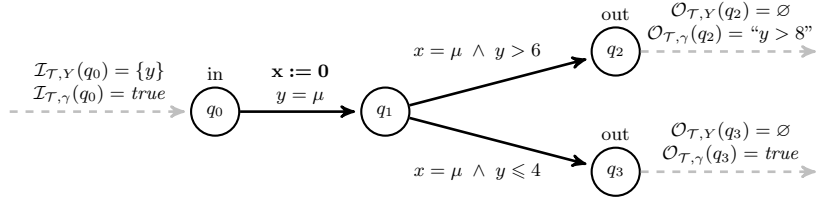

\autoref{fig:pta12short} shows an example of parametric tile $\mathcal{T}$.
For the transition from $q_0$ to $q_1$ to fire, clock $y$, reset to 0 upon entering $\mathcal{T}$ as imposed by $\mathcal{I}_{\mathcal{T}, Y}(q_0) = \{ y \}$, must have a value equal to $\mu$.
Next, for the transition from $q_1$ to $q_2$ to fire, clock $x$ must be equal to $\mu$ (it was reset when entering $q_1$), while clock $y$, thanks to the constraint on clock $x$, must satisfy $6 < y = 2\mu$, implying $3 < \mu$ holds.
A similar argument applies for the transition from $q_1$ to $q_3$, 
implying $0 < \mu \leqslant 2$ holds.
{Despite enforcing the interval $(3, +\infty)$ upon reaching $q_2$, constraint $\mathcal{O}_{\mathcal{T}, \gamma}(q_2)$ requires that $y > 8$ must hold if $\mathcal{T}$ is connected via the transition exiting $q_2$}.
%
%

A parametric tile $\mathcal{T}$ is \emph{elementary} if the connection in sequence of two instances $\mathcal{T}_1$, $\mathcal{T}_2$ of $\mathcal{T}$, i.e., $\Upsilon \subseteq \mathit{Out}_{\mathcal{T}_1} \times \mathit{In}_{\mathcal{T}_2} \times \Gamma(X_{\Theta}) \times \Sigma \times 2^{X_{\Theta}} $, for $\Theta = \{\mathcal{T}_1, \mathcal{T}_2\}$, can generate any interval in which to constrain $\mu$.
%
%
%
The tile $\mathcal{T}$ of \autoref{fig:tilexample}
is elementary.
\autoref{fig:pta12} shows a tile obtained by connecting in sequence two instances of $\mathcal{T}$: $\mathcal{T}_1$ with $n = 4$ and $\sim \; = \; >$, and $\mathcal{T}_2$ with $n = 8$ and $\sim \; = \; \leqslant$.

A \emph{Parametric TTA} $\mathcal{A}$ (shortened as PTTA) is a TTA containing at least one parametric tile.
Tile transition relations are assumed to be non-resetting test.
{From now on, we use the term \emph{tile} to refer to a parametric tile.}

{We require that, for every tile $\mathcal{T}$ of a PTTA $\mathcal{A}$, the compatibility function $\mathcal{I}_{\mathcal{T}, \gamma}$ guarantees that each parameter set $\delta_q$ and $\delta_{\mathcal{T}}$ remains unchanged for all clock values satisfying the guards specified by  $\mathcal{I}_{\mathcal{T}, \gamma}$.}
This condition and the assume-guarantee constraint of \autoref{def:tilesequence} together ensure that, in a run, parameter sets do not depend on the values of the clocks upon entering a tile. Therefore, when verifying the non-emptiness of $\mathcal{A}$ using its Underlying Weighted Automaton, the actual values of the clocks upon entering or exiting a tile can be ignored.
%

\newcommand{\creset}{\textcolor{black}}
\newcommand{\cguard}{\textcolor{black}}
\begin{figure}[b]
\begin{center}
    \begin{tikzpicture}[
	->,
	>=stealth',
	shorten >=2pt, 
	auto,
	scale=0.8,
    transform shape, 
    align=center,
    state/.style={thick, circle, draw, minimum size=0.8cm}
    ] 
    
	\node[state, label=above:{in}] (s0) {$q_0$};
	\node[state, right = 2cm of s0] (s1) {$q_1$}; 
	\node[state, label=above:{out}, right = 3.5cm of s1] (s2) {$q_2$};

    \node[state, scale = 0.3, draw=white, left = 3cm of s0] (s00) {$ $}; 
	\node[state, scale = 0.3, draw=white, right = 3cm of s2] (s01) {$ $};


		    
	\draw [line width=0.35mm]
	(s0) edge node{\creset{$\mathbf{x \boldsymbol{:=} 0}$} \\ \cguard{$y = \mu$}} (s1)
	
    (s1) edge node{\cguard{$x = \mu \; \land \; y \sim n$}} (s2)
    
    (s00) edge[dashed, lightgray] node[text=black] {
    $\mathcal{I}_{\mathcal{T}, Y}(q_0) = \{ y \}$
    \\
    $\mathcal{I}_{\mathcal{T}, \gamma}(q_0) = \mathit{true}$
    } (s0)

    (s2) edge[dashed, lightgray] node[text=black] {
    $\mathcal{O}_{\mathcal{T}, Y}(q_2) = \varnothing$
    \\
    $\mathcal{O}_{\mathcal{T}, \gamma}(q_2) = \mathit{true}$
    } (s01)
    ;
    \end{tikzpicture}
    \end{center}
    \caption{Elementary tile $\mathcal{T}$ forcing the interval $\mu \sim \! \frac{n}{2}$, where $\sim \; \in \{ < , \leqslant, \geqslant, >\}$.
    Here, $n$ is a tile-specific integer constant which must be substituted to obtain the desired interval. $\mathcal{T}$ does not assume any constraint on clock $x$ and resets clock $y$ upon entrance.}
    \label{fig:tilexample}
\end{figure}
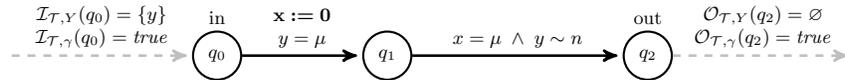
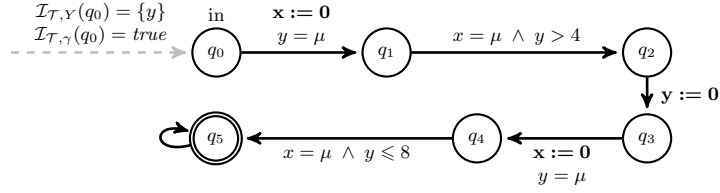
\begin{figure}[t]
\begin{center}
\begin{tikzpicture}[
	->,
	>=stealth',
	shorten >=2pt, 
	auto,
	scale=0.8,
    transform shape, 
    align=center,
    state/.style={thick, circle, draw, minimum size=0.8cm}
    ] 
    
	\node[state, label=above:{in}] (s0) {$q_0$};
	\node[state, right = 2cm of s0] (s1) {$q_1$}; 
	\node[state, right = 3.5cm of s1] (s2) {$q_2$}; 
	\node[state, below = 0.6cm of s2] (s3) {$q_3$};
	\node[state, left = 2cm of s3] (s4) {$q_4$};
	\node[state, label=, accepting, left = 3.5cm of s4] (s5) {$q_5$};

    \node[state, scale = 0.3, draw=white, left = 3cm of s0] (s00) {$ $}; 


		    
	\draw [line width=0.35mm]
	(s0) edge node{$\mathbf{x \boldsymbol{:=} 0}$ \\ $y = \mu$} (s1)
	
    (s1) edge node{$x = \mu \; \land \; y > 4$} (s2)
	
    (s2) edge node[xshift=3pt]{$\mathbf{y \boldsymbol{:=} 0}$} (s3)
	
    (s3) edge node{$\mathbf{x \boldsymbol{:=} 0}$ \\ $y = \mu$} (s4)

    (s4) edge node{$x = \mu \; \land \; y \leqslant 8$} (s5)
	
    (s5) edge[loop left] (s5)

    (s00) edge[dashed, lightgray] node[text=black] {
    $\mathcal{I}_{\mathcal{T}, Y}(q_0) = \{ y \}$
    \\
    $\mathcal{I}_{\mathcal{T}, \gamma}(q_0) = \mathit{true}$
    } (s0)
    ;
\end{tikzpicture}
\end{center}
\caption{Accepting tile $\mathcal{T}$ (obtained combining two elementary tiles in sequence) forcing the interval $(2, 4]$, hence $\delta_{\mathcal{T}} = \{(2, 4]\}$.}
\label{fig:pta12}
\end{figure}

\subsubsection{Oracles for PTTA}
We now introduce an \emph{ad hoc} weight function for PTTA over the $W^{k-\text{bit}} = (\mathcal{B}_k, \mathit{OR}, \mathit{AND}, 0^k,1^k)$ semiring:
\begin{definition}[PTTA weight function]
\label{def:weight-of-TTA}
{Let $\mathcal{A}$ be a PTTA with tile set $\Theta$, $C$ be the maximum constant appearing in the guards of $\mathcal{A}$, and $k=8C+2$.
A \emph{weight function} for $\mathcal{A}$ over $W^{k-\text{bit}}$ is a mapping $\thetaWeight{} : (\mathbb{B} \cup \bigcup_{\mathcal{T} \in (\Theta \setminus \mathbb{B})} \mathit{Out}_{\mathcal{T}}) \to \mathcal{B}_k$.}
\end{definition}

In fact, according to~\autoref{thm:parametric}, a PnrtTA can be studied by picking one or more of the following $k=8C+2$ intervals: 
\begin{equation}
\label{eq:intervals}
\{0\},
(0,\frac{1}{2}),
\{\frac{1}{2}\},
(\frac{1}{2},1),
\{1\},
(1,\frac{3}{2}),
\dots,
(\frac{4C-1}{2},2C),
\{2C\},
(2C,+\infty).
\end{equation}
We can define a bijection between $\mathcal{B}_k$ and the powerset of all intervals shown in \autoref{eq:intervals} by
uniquely associating each word in $\{0,1\}^k$ with a set of intervals: a word $w \in \{0,1\}^k$ has a $1$ in position $i$ if, and only if, the $i$-th interval is in the set.
For instance, the word $01^30^{k-4}$ maps to the interval $(0,1) = (0,\frac{1}{2}) \cup \{\frac{1}{2}\} \cup (\frac{1}{2},1)$; the word $0^{6}1^{k-6}$ maps to $[\frac{3}{2}, + \infty)$, and the word $01^{8}0^{k-9}$ maps to $(0,2]$.

\begin{definition}[Productivity]\label{def:productivity}
Let  $\mathcal{A} = (\Theta, \mathcal{T}_{0},\Sigma, \mathbb{B}, X_{\Theta}, \Upsilon)$ be a PTTA and $\thetaWeight{}$ be a weight function as in \autoref{def:weight-of-TTA}.
A tile $\mathcal{T}\in (\Theta \setminus \mathbb{B})$ is \emph{productive} for $\thetaWeight{}$ if, for each $q \in \mathit{Out}_{\mathcal{T}}$, $\thetaWeight{}(q)$ is uniquely associated with $\delta_q$ and $\delta_q$ is not empty. 
Analogously, a tile $\mathcal{T} \in \mathbb{B}$ is \emph{productive} for $\thetaWeight{}$ if 
$\thetaWeight{}(\mathcal{T})$ is uniquely associated with $\delta_\mathcal{T}$ and $\delta_\mathcal{T}$ is not empty.
\end{definition}
%
%
%
\begin{theorem}
 \label{thm:productivity}
 {Let $\mathcal{A} = (\Theta, \mathcal{T}_{0},\Sigma, \mathbb{B}, X_{\Theta}, \Upsilon)$ be a PTTA and $\thetaWeight{}$ be a weight function as in \autoref{def:weight-of-TTA}.
   If every tile in $\Theta$ is productive for $\thetaWeight{}$, checking the non-emptiness of $\mathcal{A}$ is equivalent to checking the existence of words with non-zero weight in $\uwa{}$.
   In other words, $\mathcal{L}(\mathcal{A})$ is non-empty if, and only if, 
   there exists $y \in \Sigma^*$ and a path $P$ in $\uwa{}$ with a non-zero weight in $\mathcal{B}_k$ such that $l(P) = y$.}
\end{theorem}

%

\ite{}{\Red{The proof of \autoref{thm:productivity} can be found in [XX]}.}
By analyzing each tile individually, a weight can be assigned to ensure its productivity, enabling the application of the oracle verification procedure outlined in 
\autoref{alg:generalOracleTesting}.
On line 2, a PTTA $\mathcal{A}$ is randomly generated by using a set $\Theta$ of productive tiles, weighted over the semiring $W^{k-bit}$. 
Next, on line 3, the tool $T$ under test is run with $\Aflat{}$ as input, producing a pair of results: \emph{isEmpty}, a Boolean indicating if $T$ detects an empty language, and \emph{witness}, a path leading to a B\"uchi acceptance condition in $\Aflat{}$, if it exists.
If the condition on line 4 holds, a path $P$ in $\uwa{}$ corresponding to \emph{witness} is constructed: $w(P)$ is then checked to actually be non-zero (regarding $W^{k-bit}$): if non-zero, then $T$ passes the test, else it fails.
If the condition on line 4 does not hold, the existence of words with non-zero weight (regarding $W^{k-bit}$) in $\uwa{}$ is verified: if none exist, then $T$ passes the test, else it fails.

The decidability and complexity of checking the existence of words with a non-zero weight in a Weighted Automaton depend on the chosen semiring. 
The problem is decidable for the Boolean words semiring $W^{k-\text{bit}}$ and its complexity is in polynomial-time when the constant $k$ is coded in unary, since it can be reduced to a generalization of the Floyd-Warshall algorithm for solving the shortest distance problem in a complete semiring~\cite{Mohri2009}.  


\begin{algorithm}[t]
\caption{Testing a PTA tool using oracles}
\label{alg:generalOracleTesting}

    \KwData{$\Theta$: a set of productive tiles weighted over $W^{k-\text{bit}}$, $T$: a tool to test.}
    \KwResult{\emph{true} if $T$ behaves correctly, \emph{false} otherwise.}
    
\Begin() {
    $\mathcal{A} \gets$ \texttt{GenerateRandomPTTA}$(\Theta)$\;

    $\langle \mathit{isEmpty}, \mathit{witness} \rangle \gets$ \texttt{RunTool}$(T, \Aflat{})$\;

    \If {$\mathit{isEmpty} = \mathit{false} \land \mathit{witness} \neq \varnothing$} {
        $P \gets$ path in $\uwa$ corresponding to \emph{witness}\;
        \Return {$w(P) \neq 0_{\mathcal{B}_k}$}\; } 
    \Return \texttt{NonZeroWords}$(\uwa{})$\;
    }
\end{algorithm}

\section{Experimental results}
\label{sec:sec6}\label{sec:tester}
This section explains how oracle generation is implemented in practice in \toolname{} and presents the results of its experimental evaluation.

\subsection{Forecasting Parameter Values with \toolname{}}

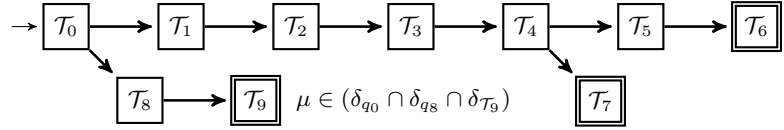
\begin{figure}[tb]
\begin{center}
\begin{tikzpicture}[
	->,
	>=stealth',
	shorten >=2pt, 
	auto,
	scale=1,
    transform shape, 
    align=center,
    state/.style={thick, rectangle, draw, minimum size=0.6cm}
    ] 
    
	\node[state, initial, initial text={}] (s0) {$\mathcal{T}_{0}$};
    
	\node[state, right = 0.9cm of s0] (s1) {$\mathcal{T}_{1}$}; 
    
	\node[state, below right = 0.5cm of s0] (s2) {$\mathcal{T}_{8}$}; 
	
	\node[state, accepting, right = 0.9cm of s2] (s9) {$\mathcal{T}_{9}$};
	
	\node[state, right = 0.9cm of s1] (s3) {$\mathcal{T}_{2}$}; 
    
	\node[state, right = 0.9cm of s3] (s4) {$\mathcal{T}_{3}$};
    
    \node[state, right = 0.9cm of s4] (s5) {$\mathcal{T}_{4}$};
    
	\node[state, right = 0.9cm of s5] (s6) {$\mathcal{T}_{5}$};
    
	\node[state, accepting, below right = 0.5cm of s5] (s7) {$\mathcal{T}_{7}$};
	
	\node[state, accepting, right = 0.9cm of s6] (s8) {$\mathcal{T}_{6}$};
	
	\node[right= 0.1cm of s9] {
    $\mu \in (\delta_{q_0}
              \cap
              \delta_{q_8}
              \cap
              \delta_{\mathcal{T}_9})$
    };
	
	\draw [line width=0.35mm]
	(s0) edge (s1)
	(s1) edge (s3)
	(s3) edge (s4)
	(s4) edge (s5)
	(s5) edge (s6)
	(s5) edge (s7)
	(s0) edge (s2)
    (s6) edge (s8)
	(s2) edge (s9);
\end{tikzpicture}
\end{center}
\caption[]{Example of a binary tree-like PTTA. For simplicity, compatibility functions are omitted.
A possible valid set of intervals for $\mu$ is shown on the right of tile $\mathcal{T}_9$. 
It is computed as the intersection of the parameter sets of the tiles traversed to reach $\mathcal{T}_9$.
Here, $q_0 \in \mathit{Out_{\mathcal{T}_0}}$ (from which the transition from $\mathcal{T}_0$ to $\mathcal{T}_8$ originates) and $q_8 \in \mathit{Out_{\mathcal{T}_8}}$.}
\label{fig:treetiledta}
\end{figure}

\toolname{} allows users to generate PTTA (manually or automatically) as a composition of tiles, and subsequently test their emptiness.
%
%
At the time of writing, \toolname{} is still a prototype; in particular, tiles 
must have only one input location and up to two output locations.
Furthermore, the input location of a tile must have at most one incoming transition.
For these reasons, the PTTA considered by the current version of \toolname{} have a \emph{binary tree} structure, as shown for instance in~\autoref{fig:treetiledta}. 
{The binary tree structure cannot model all real-world systems; still, it acts as a reference scalability performance indicator: if performance is poor on binary tree-like PTTA, it will be poor on generic PTTA as well.}

By \autoref{thm:productivity}, \toolname{} can predict the interval in which the parameter value may fall (if any exists) before performing the actual emptiness test, thus producing a test oracle.
{For practical reasons, TABEC operates using unions and intersections of intervals, which correspond to the $\mathit{OR}$ and $\mathit{AND}$ operations over Boolean words in the semiring $W^{k\text{-bit}}$ performed by Underlying Weighted Automata.}
In the case of binary trees, each path $P$ from $\mathcal{T}_0$ to an accepting tile $\mathcal{T}$ is a sequence of tiles: the set of intervals in which the parameter value may fall upon reaching $\mathcal{T}$  over $P$ is obtained by intersecting the parameter sets of the tiles in $P$, considering only those related to the output locations from which the transitions in $P$ originate, along with the parameter set of $\mathcal{T}$.
The complete set of admissible intervals is obtained by repeating this computation for each path from $\mathcal{T}_0$ to an accepting tile.
\autoref{fig:treetiledta} provides an example of this computation when considering the path $P = \mathcal{T}_0\mathcal{T}_8\mathcal{T}_9$.
The set of intervals obtained upon reaching tile $\mathcal{T}_9$ from $\mathcal{T}_0$ in $P$ is computed as
$\delta_{q_0}
\cap
\delta_{q_8}
\cap
\delta_{\mathcal{T}_9}$, where $q_0 \in \mathit{Out_{\mathcal{T}_0}}$ is the origin of the transition from $\mathcal{T}_0$ to $\mathcal{T}_8$ and $q_8 \in \mathit{Out_{\mathcal{T}_8}}$.

{To validate our theoretical results, extensive testing was performed on a Linux machine (AMD EPYC 7282 CPU, 16 cores, 32 threads, 2.8-3.2 GHz).}
For brevity, in this section we refer to ``TA'', though in our experiments \toolname{} analyzed randomly generated PTTA. 
We also define the \emph{size} of a TA as the sum of its locations and transitions.

\subsection{Scalability of \toolname{}}
\label{sec:sec6-1}
A test suite evaluating the scalability of \toolname{} was initially conducted.
These tests were performed on both \hyperref[alg:mainalgo]{EmpCheck} and EmpCheckFast. 
Timeouts of 62 and 122 minutes were set to stop test execution when exceeding this time limit.
A total of 367 tests were performed.
The second row of \autoref{table:algoscalab} contains the biggest TA size obtained for both versions of the algorithm run by setting the aforementioned timeouts.
\hyperref[alg:mainalgo]{EmpCheck} revealed to be slower, since it was not able to check a larger TA before reaching the timeouts.
{This can be inferred from the first row of \autoref{table:algoscalab} showing that, considering the same timeouts, \hyperref[alg:mainalgo]{EmpCheck} executed fewer tests than EmpCheckFast}.
The correctness of the results provided by tChecker was evaluated using \toolname{}'s oracle capabilities. 129 out of 367 tests admitted a B\"uchi acceptance condition. 
For these tests, the parameter value interval was computed by the verification algorithm with 100\% accuracy.

\definecolor{bluepoli}{cmyk}{0,0,0,0.2}
\newcolumntype{C}{>{}c}
\newcolumntype{L}{>{}l}
\begin{table}[t]
\centering 
\captionsetup{skip=10pt}
\footnotesize
\resizebox{0.9\textwidth}{!}{
\renewcommand{\arraystretch}{1.27}
    \begin{tabular}{| L  C  C  C  C |}
    \hline
    \rowcolor{bluepoli} 
    \fontseries{b}\selectfont
        \textbf{Metric} & 
        \begin{tabular}{@{}c@{}}
            \textbf{\hyperref[alg:mainalgo]{EmpCheck}}
        \end{tabular} & 
        \begin{tabular}{@{}c@{}}
		    \textbf{\hyperref[alg:mainalgo]{EmpCheck}}
		\end{tabular} & 
		\begin{tabular}{@{}c@{}}      
            \textbf{EmpCheckFast}
        \end{tabular} & 
        \begin{tabular}{@{}c@{}}
            \textbf{EmpCheckFast}
        \end{tabular} \\
        \hline \hline
        \begin{tabular}{@{}l@{}}
            \#Tests
        \end{tabular} & 79 & 103 & 81 & 104 \\
        \hline
        \begin{tabular}{@{}l@{}}
            MaxSize 
        \end{tabular} & 869 & 1172 & 911 & 1199 \\
        \hline
        \begin{tabular}{@{}l@{}}
            \#NonEmpty
        \end{tabular} & 28 & 38 & 25 & 38 \\
        \hline
        \begin{tabular}{@{}l@{}}
            \#Empty
        \end{tabular} & 51 & 65 & 56 & 66 \\
        \hline
        \begin{tabular}{@{}l@{}}
            Accuracy
        \end{tabular} & 100\% & 100\% & 100\%  & 100\%\\
        \hline
        \begin{tabular}{@{}l@{}}
            Timeout[min]
        \end{tabular} & 62 & 122 & 62 & 122\\
        \hline
    \end{tabular}
    }
    \caption{Scalability results of \toolname{} considering \hyperref[alg:mainalgo]{EmpCheck} and EmpCheckFast.}
    \label{table:algoscalab}
\end{table}

\subsection{Resource utilization testing}
\label{subsec:restest}
A second test suite was conducted to measure tChecker resource utilization for each call made to the tool, i.e., each call to \emph{checkEmp()} of \autoref{fig:component}, using the EmpCheckFast algorithm and focusing on computation time and memory. 
A total of 18 tests of increasing complexity were generated, iteratively incrementing TA size at steps of 40 locations and 80 transitions each. 
The maximum constant appearing in the generated TA was 10.
Measurements are shown in \autoref{fig:tcktiming}, where it is clear that the overall tChecker execution time does not increase linearly with the size of TA. The figure contains mean measurements derived by averaging results obtained during each test execution.
In particular, each test is weighted by the number of times tChecker was invoked by EmpCheckFast within that test, in order to obtain a valid parameter value.
{The bars above individual columns represent the highest measured value for each specific test}.
\pgfplotsset{compat=1.16}

\usetikzlibrary{patterns}

\begin{figure*}[h]
\centering
\begin{tikzpicture}
  \begin{axis}[
    width=0.92\textwidth,height=65mm,
    x tick label style = {font = \small, align = center, rotate = 60},
    xtick={1, 2, 3, 4, 5, 6, 7, 8, 9, 10, 11, 12, 13, 14, 15, 16, 17, 18},
    xticklabels={19,139,259,379,499,619,739,859,979,1099,1219,1399,1459,1579,1699,1819,1939,2059},
    ylabel={Time [seconds]},
    xlabel={TA size $[\lvert Q \rvert + \lvert T \rvert]$},
    xlabel style={yshift=0pt},
    ymajorgrids,
    major grid style={dashed, color=black!35},
    axis line style={thick},
    tick style={thick},
    legend cell align={left},
    ytick pos=left, 
    legend style={
      at={(0,1.2)},
      anchor=north west,
    },
    ybar=1pt,
    ymin=-9.5,
    bar width=5pt,
    error bars/y dir=plus, 
    error bars/y explicit, 
    error bars/error bar style={
      thick, 
      draw=green!70!black 
    }
  ]
    \addplot [
    bar shift=-3pt,
    draw=blue, 
  fill=blue!30 
  ] coordinates {
        (01,0.004) +- (0, 0)
        (02,0.0848095) +- (0, 0.0801905)
        (03,0.523167) +- (0, 0.007833)
        (04,1.53069) +- (0, 0.01331)
        (05,3.4211) +- (0, 0.0229)
        (06,6.64817) +- (0, 0.07383)
        (07,11.3689) +- (0, 0.5291)
        (08,17.501) +- (0, 0.101)
        (09,25.927) +- (0, 0.273)
        (10,37.1728) +- (0, 0.3922)
        (11, 49.3869) +- (0, 0.3751)
        (12,66.3315) +- (0, 0.4045)
        (13,89.0118) +- (0, 1.3012)
        (14,109.905) +- (0, 0.657)
        (15,136.941) +- (0, 1.163)
        (16, 169.516) +- (0, 1.509)
        (17,203.188) +- (0, 4.817)
        (18,243.948) +- (0, 1.147)};
    \legend{tChecker time [seconds]}
  \end{axis}

  \begin{axis}[
    width=0.92\textwidth,height=65mm,
    x tick label style = {font = \small, align = center, rotate = 60},
    xtick={1, 2, 3, 4, 5, 6, 7, 8, 9, 10, 11, 12, 13, 14, 15, 16, 17, 18},
    xticklabels={},
    ylabel={Memory [KBytes]},
    ymajorgrids,
    major grid style={solid, color=black!35},
    axis line style={},
    tick style={thick},
    axis y line*=right,
    ylabel style={yshift=0pt},
    legend cell align={left},
    legend style={
      at={(1,1.2)},
      anchor=north east,
    },
    ybar=1pt,
    bar width=5pt,
    error bars/y dir=plus, 
    error bars/y explicit, 
    error bars/error bar style={
      thick, 
      draw=green!70!black 
    }
  ]
    \addplot [
    pattern=north east lines,
    pattern color=red,
    preaction={fill=red!20},
    bar shift=3pt,
    draw=red 
    ] coordinates {
        (01,7.524) +- (7.52,0)
        (02,8.19381) +- (8.19381, 11.80419)
        (03,8.21324) +- (8.21324, 0.12676)
        (04,8.53076) +- (8.53076, 0.11724)
        (05,8.81829) +- (8.81829, 0.12171)
        (06,9.27457) +- (9.27457, 0.21343)
        (07,12.0237) +- (12.0237, 47.0523)
        (08,12.7051) +- (12.7051, 0.1909) 
        (09,14.7567) +-  (14.7567, 0.1513)
        (10,16.9383) +-  (16.9383, 0.1857)
        (11,19.3845) +-   (19.3845, 0.1355)
        (12,22.0354) +-  (22.0354, 0.1246)
        (13,24.9627) +- (24.9627, 0.1773) 
        (14,28.0529) +-  (28.0529, 0.1351)
        (15,31.4702) +- (31.4702, 0.1938)
        (16, 34.9732) +- (34.9732, 0.0828)
        (17,41.0002) +-  (0, 92.6118)
        (18,42.9758) +-  (42.9758, 0.1162)};
    \legend{tChecker memory [KBytes]}

  \end{axis}
\end{tikzpicture}
\caption[Results for tChecker]{Plot containing mean time and memory used by tChecker during tests.}
\label{fig:tcktiming}
\end{figure*}
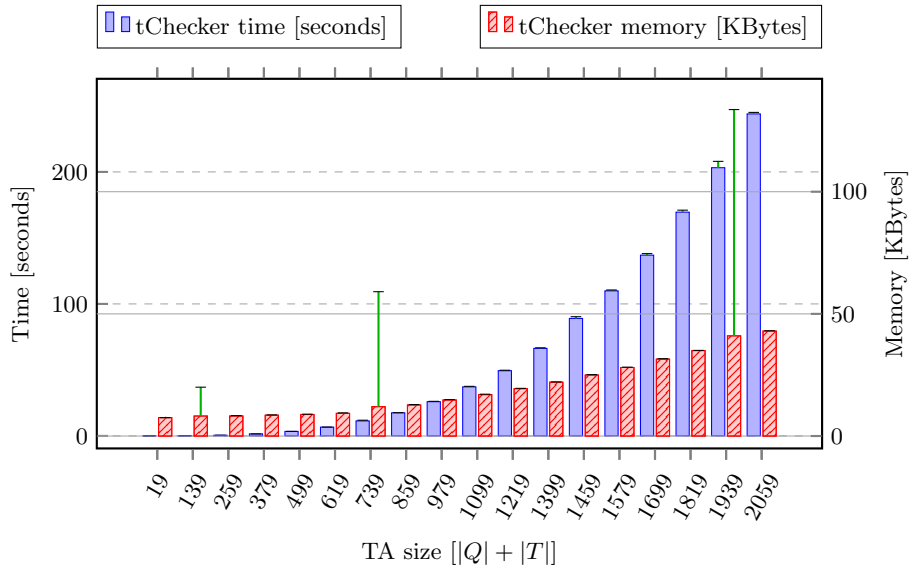 

\section{Oracle Generation for Priced Timed Automata}
\label{sec:pta}
In this section we extend the oracle generation framework introduced in \autoref{sec:forecasting} to another formalism, namely \emph{Priced Timed Automata} (shortened as PriTA)~\cite{pta-article}.

\subsubsection{Priced Timed Automata}
A PriTA $\mathcal{A}$ is obtained by adding a \emph{cost function} $c : (Q \cup T) \rightarrow \mathbb{N}$ to \autoref{def:tadef} of TA, assigning costs to locations and transitions.
The overall cost of a location $q \in Q$ equals the value specified by $c(q)$ multiplied by the total time $\mathcal{A}$ stays in $q$;
for a transition $t \in T$, it corresponds to the value specified by $c(t)$. 
For convenience, we assume that given a pair of locations $ \langle q,q'\rangle$, each transition from $q$ to $q'$ has the same cost, denoted by $c(\langle q ,q'\rangle)$.
As customary for PriTA, we refer to the set $B$ of accepting locations as the set of \emph{goal} locations.
Let 
$
\rho_n = 
(q_{0}, v_{0})
(q_{1}, v_{1})
\dots
(q_{n}, v_{n})
$
be a finite run
of $\mathcal{A}$ over a timed word $(\sigma, \tau)$; its total cost, up to configuration $n$
($n \geqslant 1$), is computed as:
\begin{equation*}
\mathit{Cost}(\rho_{n}) = 
\sum^{n-1}_{k = 0} 
c(q_{k}) \cdot (\tau_{k+1} - \tau_k) + c(\langle q_k,q_{k+1}\rangle)
\text{.}
\end{equation*}

Finite runs are used to solve the \emph{minimum-cost reachability problem}, i.e., finding the minimum cost (often, a lower bound on the cost), denoted by \emph{mc$(B)$}, to reach the set $B \subseteq Q$ of goal locations:
\begin{equation*}
\text{\emph{mc}}(B) = 
\inf \{\text{\emph{Cost}}(\rho_n)
~
\vert
~
\rho_n = 
(q_{0}, v_{0})
(q_{1}, v_{1})
\dots
(q_{n}, v_{n}) 
~
\land
~
q_{n} \in B \} \text{.}
\end{equation*}

\begin{figure}[b]
\begin{center}
\begin{tikzpicture}[
	->,
	>=stealth',
	shorten >=2pt, 
	auto,
    transform shape, 
    align=center,
    scale=0.8,
    state/.style={thick, circle, draw, minimum size=0.8cm}
    ] 
    
	\node[state, label=above:{$c(q_0)=1$}, initial, initial text={}]         (s0) {$q_0$};
	\node[state, label=above:{$c(q_1)=2$}, right = 3cm of s0] (s1) {$q_1$};
	\node[state, label=above:{$c(q_2)=1$}, right = 3cm of s1] (s2) {$q_2$};
	\node[state, label=above:{$c(q_3)=1$}, right = 3cm of s2, accepting] (s3) {$q_3$};

	\draw [line width=0.35mm]
	(s0) edge node{$\mathbf{x \boldsymbol{:=} 0}$ \\ $x > 0 \land x = 2$} node[below]{$c(t_0) = 4$}(s1)
	(s1) edge node{$\mathbf{x \boldsymbol{:=} 0}$ \\ $x > 0 \land x = 3$} node[below]{$c(t_1) = 0$}(s2)
	(s2) edge node{$\mathbf{x \boldsymbol{:=} 0}$ \\ $x > 0 \land x = 1$} node[below]{$c(t_2) = 1$}(s3)
	;
\end{tikzpicture}
\end{center}

\caption[PriTA Example]{Example of PriTA.
The cost function $c$ is shown on each location and transition.}
\label{fig:ptaExample}
\end{figure}
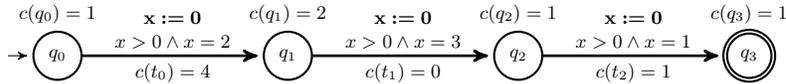

An example of PriTA $\mathcal{A}$ is given in \autoref{fig:ptaExample}. 
We assume that, before the execution starts from location $q_0$, the total cost of $\mathcal{A}$, denoted by $\mathit{Cost}_{\mathcal{A}}$, is equal to 0.
The guard on transition $t_0$ imposes that exactly 2 time units must elapse for $t_0$ to fire.
Thus, when entering location $q_1$, $\mathit{Cost}_{\mathcal{A}}$ has the following value: $\mathit{Cost}_{\mathcal{A}} = c(q_0) \cdot 2 + c(t_0) = 1 \cdot 2 + 4 = 6$.
Applying the same reasoning to the run
$
\rho_3 = 
(q_{0}, v_{0})
(q_{1}, v_{1})
(q_{2}, v_{2})
(q_{3}, v_{3})
$
gives a total cost of 
$
\mathit{Cost}_{\mathcal{A}}(\rho_3) = 
c(q_0) \cdot 2 + c(t_0) + 
c(q_1) \cdot 3 + c(t_1) +
c(q_2) \cdot 1 + c(t_2) = 
14$.


\subsubsection{Tiled PriTA}

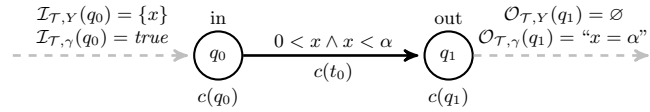
\begin{figure}[tb]
\begin{center}
\begin{tikzpicture}[
	->,
	>=stealth',
	shorten >=2pt, 
	auto,
    transform shape,
    align=center,
    scale=0.8,
    state/.style={thick, circle, draw, minimum size=0.8cm}
    ] 

     \node[state, label=below:$c(q_0)$, label=above:in] (q0) {$q_0$};
     
    \node[state, label=below:$c(q_1)$, label=above:out, right=3cm of q0] (q1) {$q_1$};

    \node[state, scale = 0.3, draw=white, left = 3cm of q0] (s00) {$ $}; 
	\node[state, scale = 0.3, draw=white, right = 3cm of q1] (s01) {$ $};




	\draw [line width=0.35mm]
	(q0) edge node{$0 < x \land x < \alpha$} node[below]{$c(t_0)$}(q1)
	
    (s00) edge[dashed, lightgray] node[text=black] {
    $\mathcal{I}_{\mathcal{T}, Y}(q_0) = \{ x \}$
    \\
    $\mathcal{I}_{\mathcal{T}, \gamma}(q_0) = \mathit{true}$
    } (q0)

    (q1) edge[dashed, lightgray] node[text=black] {
    $\mathcal{O}_{\mathcal{T}, Y}(q_1) = \varnothing$
    \\
    $\mathcal{O}_{\mathcal{T}, \gamma}(q_1) = \text{``}x = \alpha\text{''}$
    } (s01)

    ;
\end{tikzpicture}
\end{center}

\caption[PriTA tile]{Example of a PriTA tile, where
$\alpha \in \mathbb{N}_{> 0}$ is a tile-specific constant.
The function $c$ assigns costs to the locations $q_0$ and $q_1$ and to the transition $t_0$.
}
\label{fig:ptaTile}
\end{figure}

A cost function $c$ can be added to \autoref{def:tiledef} of tiles, obtaining so-called \emph{PriTA tiles}.
\autoref{fig:ptaTile} provides an example of a PriTA tile $\mathcal{T}$.
Since the compatibility function  $\mathcal{I}_{\mathcal{T}, Y}$ resets clock $x$  upon entering location $q_0$ and time is strictly monotonic, the guard on transition $t_0$ requires a value of $0 < x < \alpha$ time units to elapse before exiting $q_0$.
The compatibility function  $\mathcal{O}_{\mathcal{T}, \gamma}$ imposes a constraint requiring that clock $x$ is exactly equal to $\alpha$ when exiting $\mathcal{T}$.
%
The lower bound  on the cost accrued upon exiting $\mathcal{T}$ from location $q_1$ is equal to $\mathit{Cost}_{\mathit{out}}$ = $\alpha \cdot p + c(t_0)$, where $p = \min(c(q_0), c(q_1))$.

In general, given a tile $\mathcal{T}$, the value $\mathit{Cost}_{\mathit{out}}\in \mathbb{N}$ is computed as the lower bound on the cost required to traverse $\mathcal{T}$ and exiting from an output location $q \in \mathit{Out}_{\mathcal{T}}$. 
This value is called the \emph{weight} of the tile at the output location $q$.



A \emph{Tiled PriTA} $\mathcal{A}$
is a TTA $\mathcal{A} = (\Theta, \mathcal{T}_{0},\Sigma, \mathbb{B}, X_{\Theta}, \Upsilon)$ as in \autoref{def:tiledTA}, where $\Theta$ contains only PriTA tiles.
%
%
We now derive an oracle for testing model checkers solving the min-cost reachability problem. 
Let $\thetaWeight{} : (\mathbb{B} \cup \bigcup_{\mathcal{T} \in (\Theta \setminus \mathbb{B})} \mathit{Out}_{\mathcal{T}}) \to \mathbb{N}$ be a weight function mapping accepting tiles and output locations of non-accepting tiles to their corresponding weight, and $w_{\Upsilon} : \Upsilon \to \mathbb{N}$ be a weight function mapping each transition of $\mathcal{A}$ to its cost. 
The definition of the Underlying Weighted Automaton of $\mathcal{A}$ over the min-plus semiring $W^{\text{price}}= (\mathbb{N}\cup \{+\infty\}, \mathit{min}, +, +\infty,0)$ can be easily obtained by adapting $w_{\Upsilon}$ to \autoref{def:underlyingWA}. 
Solving min-cost reachability for $\mathcal{A}$ is equivalent to solving the same problem for $\uwa{}$.
Indeed, also in the case of  $W^{\text{price}}$, this problem can be reduced to solving a {shortest distance problem by applying the polynomial-time algorithm of~\cite{Mohri2009}.}
\section{Conclusion and future works}
\label{sec:sec7}\label{sec:future}
{This work presented a novel theoretical framework for rigorously testing the correctness of model checkers 
solving verification problems for variations of Timed Automata.
Our framework generates  a Tiled Timed Automaton (i.e., a modular Timed Automaton) and reduces it to a Weighted Automaton used as an oracle.
A tool, called \toolname{}, was developed to validate the framework in the context of the emptiness problem by using parametric non-resetting test Timed Automata.

Empirical results underscore the robustness of our framework, providing a solid foundation for future developments and improvements.
The current implementation of \toolname{} restricts tile composition to binary trees.
However, our framework supports more general structures for tile composition (e.g., cycles between different tiles), therefore we plan to extend \toolname{} in this direction.}

We also described how to build oracles for model checkers solving the min-cost reachability problem for Priced TA.
%
%
Our framework, however, is not confined to PnrtTA and Priced TA; we plan to adapt it to other TA variants, expanding the spectrum of model checkers and their corresponding solvable problems eligible for oracle testing.
%
This would require a suitable definition of tiles, along with a careful identification of the appropriate semiring for each problem at hand.
%
%
%
%

\bibliographystyle{splncs04}
\bibliography{bibliography}

\ite{
\newpage 

\appendix
\newcommand{\tile}{\mathcal{T}}
\newcommand{\acctile}{\tile_\mathit{acc}}
\newcommand{\aut}{\mathcal{A}}
\newcommand{\autw}{\mathcal{W}}
\newcommand{\accset}{\mathbb{B}}
\newcommand{\pathLength}{\lvert P \rvert}
\newcommand{\patha}{P_{\mathcal{A}}}
\newcommand{\pathw}{P_{\mathcal{W}}}
\newcommand{\pathaLength}{\lvert \patha \rvert}
\newcommand{\pathwLength}{\lvert \pathw \rvert}
\newcommand{\langaut}{\mathcal{L}(\mathcal{A})}
\newcommand{\prefw}{W}
\newcommand{\outT}{w_T(\tile_0, y_1, \tile_1)}
\newcommand{\TsubB}{(\Theta \setminus \accset)}
\newcommand{\QsubF}{(Q \setminus F)}
\newcommand{\pathapLength}{\lvert \patha' \rvert}
\newcommand{\pathwpLength}{\lvert \pathw' \rvert}

\setcounter{theorem}{1}
\section{Proof of \autoref{thm:productivity}}
\label{app:proof_of_theorem_2}
\begin{theorem}
 {Let $\mathcal{A} = (\Theta, \mathcal{T}_{0},\Sigma, \mathbb{B}, X_{\Theta}, \Upsilon)$ be a PTTA and $\thetaWeight{}$ be a weight function as in \autoref{def:weight-of-TTA}.
   If every tile in $\Theta$ is productive for $\thetaWeight{}$, checking the non-emptiness of $\mathcal{A}$ is equivalent to checking the existence of words with non-zero weight in $\uwa{}$.
   In other words, $\mathcal{L}(\mathcal{A})$ is non-empty if, and only if, 
   there exists $y \in \Sigma^*$ and a path $P$ in $\uwa{}$ with a non-zero weight in $\mathcal{B}_k$ such that $l(P) = y$.}
\end{theorem}

\begin{proof}
    For simplicity, in the proof we assume that $|\Sigma|=1$, hence we can ignore the input alphabet of both $\aut$ and $\uwa$.
    
    We 
    define a \emph{path} of $\aut$ of length $n>0$ as a sequence of $n$ consecutive productive tiles (according to the tile transition relation $\Upsilon$), starting from a possibly non-initial tile, and ending in an accepting tile, namely a sequence of the form 
    $\patha = \tile_{i_1}\tile_{i_2} \dots \tile_{i_n}$, 
    with each pair of tiles $\tile_{i_j},\tile_{i_{j+1}}$ connected by a tuple of  $\Upsilon$,
    such that
    $\tile_{i_1}, \tile_{i_2}, \dots, \tile_{i_{n-1}} \in \TsubB$ and $\tile_{i_n} \in \accset$.
    Notice that non-accepting tiles can be traversed more than once.
    We define the \emph{parameter set} of $\patha$  
    as the set of parameter intervals
    {generating a finite run $\rho = (q_{i_1}, v_{i_1})(q_{i_2}, v_{i_2})\dots(q_{i_n}, v_{i_n})$ over $\Aflat{}$ such that every $q_{i_j}$ is a location of tile $\tile_{i_j}$}.
    Let $\pathw = \tile_{i_1} \tile_{i_2} \dots \tile_{i_n}$ be a path of length $n$ in $\uwa$, 
    with states $\tile_{i_1}, \tile_{i_2}, \dots, \tile_{i_{n-1}} \in \QsubF$
    and $\tile_{i_n} \in F$. Let $w(\pathw)$ be its weight.
%
    
    We first claim that the non-emptiness of $\mathcal{A}$ implies the existence of words with non-zero weight in $\uwa{}$ leading to a final state in $F$.
    The proof is by induction over the length of the paths of $\aut$.
    The induction hypothesis is that, for every path $\patha = \tile_{i_1} \tile_{i_2} \dots \tile_{i_n}$ of $\aut$ of length  $n>0$, the weight of the path $\pathw = \tile_{i_1} \tile_{i_2} \dots \tile_{i_n}$ of  $\uwa$ corresponds to the parameter set of $\patha$. 
    The claim follows immediately since, if $\patha$ starts from an initial tile and its parameter set is non-empty, then the language of $\mathcal{A}$ is non-empty (because of the assume-guarantee constraint of \autoref{def:tilesequence}), and in $\uwa{}$ there is a word with non-zero weight leading to a final state in $F$.
%

    
    (\emph{Base case}) Let $n = 1$.
    Consider the path 
    $\patha = \tile_{i_1}$, with $\tile_{i_1} \in \accset$. 
    Since $\tile_{i_1}$ is productive, it has a non-empty parameter set. By \autoref{def:productivity}, its parameter set corresponds to the weight of 
    $\pathw = \tile_{i_1}$, with $\tile_{i_1} \in F$.


    (\emph{Induction}) Suppose the induction hypothesis holds for $n \in \mathbb{N}_{> 0}$.
    We show that it also holds for $n+1$.     Consider the path
    $\patha' = \tile_{i_0} \tile_{i_1}\tile_{i_2} \dots \tile_{i_n}$, where $\tile_{i_0} \in \TsubB$, having length $\pathapLength = n+1$. 
    The parameter set of $\patha'$ is the intersection  of the parameter set of the transition from tile $\tile_{i_0}$ to tile $\tile_{i_1}$ with the parameter set of the path  $\patha=\tile_{i_1}\tile_{i_2} \dots \tile_{i_n}$. 
    The weight of the path $\pathw' = \tile_{i_0} \tile_{i_1}\tile_{i_2} \dots \tile_{i_n}$ of $\uwa$, where $\tile_{i_0} \in \QsubF$, is the $\mathit{AND}$ (bit-by-bit) of {the weight of the path $\pathw = \tile_{i_1} \tile_{i_2} \dots \tile_{i_n}$}
    and of the weight of the transition from state $\tile_{i_0}$ to state  $\tile_{i_1}$. 
    By induction hypothesis, the weight of the path
    $\pathw$ of $\uwa$ corresponds to the parameter set of $\patha$. 
    The thesis follows immediately, since the parameter set of the transition from tile $\tile_{i_0}$ to tile $\tile_{i_1}$ in $\aut$ corresponds to the weight of the transition from state  $\tile_{i_0}$ to state $\tile_{i_1}$ in $\uwa$, 
    and the intersection of two parameter sets corresponds to the $\mathit{AND}$ of the two weights.
    
    We now claim that the existence of words with non-zero weight in $\uwa{}$ leading to a final state in $F$ implies the non-emptiness of $\mathcal{A}$.
   The induction hypothesis is that, for every $n>0$, every path $\pathw = \tile_{i_1} \tile_{i_2} \dots \tile_{i_n}$ of  $\uwa$ of length $n>0$  has a weight corresponding to the parameter set of the path $\patha = \tile_{i_1} \tile_{i_2} \dots \tile_{i_n}$ of $\aut$. The proof is similar to the previous case and is omitted.
   
%
%
\hfill \qed
\end{proof}
}{}

\end{document}